\documentclass[aps,pre,twocolumn,superscriptaddress,floatfix]{revtex4-1}
\usepackage{graphicx,amsmath,amssymb,verbatim,color,graphicx}
\newcommand{\be}{\begin{equation}}
\newcommand{\ee}{\end{equation}}

\begin{document}
\title{Network robustness of multiplex networks with interlayer degree correlations}
\author{Byungjoon Min}
\affiliation{Department of Physics, Korea University, Seoul 136-713, Korea}
\author{Su Do Yi}
\affiliation{Department of Physics, Sungkyunkwan University, Suwon 440-746, Korea}
\author{Kyu-Min Lee}
\affiliation{Department of Physics, Korea University, Seoul 136-713, Korea}
\author{K.-I.~Goh}
\email{kgoh@korea.ac.kr}
\affiliation{Department of Physics, Korea University, Seoul 136-713, Korea}
\date{\today}
\begin{abstract}
We study the robustness properties of multiplex networks consisting of multiple layers of distinct types of links, focusing on the role of correlations between degrees of a node in different layers.
We use generating function formalism to address various notions of the network robustness relevant to
multiplex networks such as the resilience of ordinary- and mutual connectivity 
under random or targeted node removals as well as the biconnectivity.
We found that correlated coupling can affect the structural robustness of multiplex networks 
in diverse fashion. For example, for maximally-correlated duplex networks, 
all pairs of nodes in the giant component are connected via at least two independent paths
and network structure is highly resilient to random failure.
In contrast, anti-correlated duplex networks are on one hand robust against targeted attack 
on high-degree nodes, but on the other hand they can be vulnerable to random failure.
\end{abstract}
\maketitle

\section{Introduction}
Complex network theory has successfully accounted for structural and dynamical problems of complex systems in terms of their connectivity patterns~\cite{newman_book,cohen-havlin}. Most studies on complex networks, so far, have dealt with isolated network layers~\cite{newman_book,cohen-havlin}. However, many real-world complex systems such as physical, social, biological, and infrastructural systems consist of multiple layers of networks interacting each other~\cite{padgett,verbrugge,szell,mucha,little,rosato,mucha,cardillo,buldyrev,leicht}. Recently, several studies on multiplex networks in which a node belongs to multiple network layers of distinct types of links~\cite{kmlee,gardenes,gomez} have contributed to the progress of research on multi-layer complex systems~\cite{review} along with other approaches like interdependent~\cite{gao,parshani2,buldyrev,swson} and interconnected networks~\cite{leicht,louzada}. These studies have shown that the coupling structure 
and the interactions among different layers can significantly affect percolation \cite{kmlee,leicht}, 
diffusion~\cite{gomez}, cascade of failures~\cite{buldyrev,brummitt,schneider2}, and network evolution \cite{evolution} in such networks.

For many real-world multiplex networks, network layers are correlated one another rather than combined randomly. Although there exist various forms of correlations between network layers, the interlayer degree correlation would be one of the simplest types as observed in multiplex online game social network data~\cite{szell}. In this case, a positive correlation represents that the degree of a node in one layer tends 
to be correlated with that in other layers, such that the hub in one layer also has many neighbors in the other layers. On the contrary, the hub in one layer would have few neighbors in the other layers for 
negatively correlated multiplex networks. Recently, the effect of such interlayer degree correlation was addressed for connectivity of multiplex networks~\cite{kmlee}. Furthermore, a few studies demonstrated that interdependent networks with higher interlayer degree correlation~\cite{parshani,correlation} or more assortative layers~\cite{zhou} are more robust under random damage. However, there is still lack of unified understanding of various robustness properties of multiplex networks due to the role of interlayer degree correlations.

Network robustness refers to the structural resilience of a network to external perturbations, which has been one of the most active topics in complex networks theory \cite{cohen-havlin}. The study on the network robustness aims not only for theoretical interests \cite{albert,callaway,cohen,cohen-att,holme} but also for practical applications to design more resilient structures against random breakdowns or intentional attacks \cite{valente,tanizawa,schneider}. Backup pathway between a pair of nodes is a meaningful concept of the network robustness, captured by the connection between a pair through at least two paths, 
termed biconnectivity~\cite{bicomponent,pkim}. Since a biconnected pair in networks can communicate under removal of one route, the biconnectivity can play a significant role in the network robustness.

Another widely-used measure of the network robustness is the the size of remaining giant component after removing a fraction of nodes or links, either chosen randomly or targeted with respect to their degrees \cite{albert,callaway,cohen,cohen-att,holme}. Previous studies found that the network robustness under removal of nodes (or links) depends on the connectivity patterns of networks~\cite{albert,callaway,cohen,cohen-att}.
In multiplex networks, different types of connectivity can be meaningful depending on the context with which the multiple network layers are coupled. In addition to the usual connectivity \cite{leicht,kmlee}, for example,
the so-called mutual connectivity can be significant in multiplex networks with cooperative or interdependent layers, in which case a node requires simultaneous connectivities through each and every layer for proper functioning~\cite{buldyrev}. Here, we study the impact of the interlayer degree correlation on various robustness properties of multiplex networks in terms of the biconnectivity, the connectivity, and the mutual connectivity. 

\begin{figure}
\includegraphics[width=0.8\linewidth]{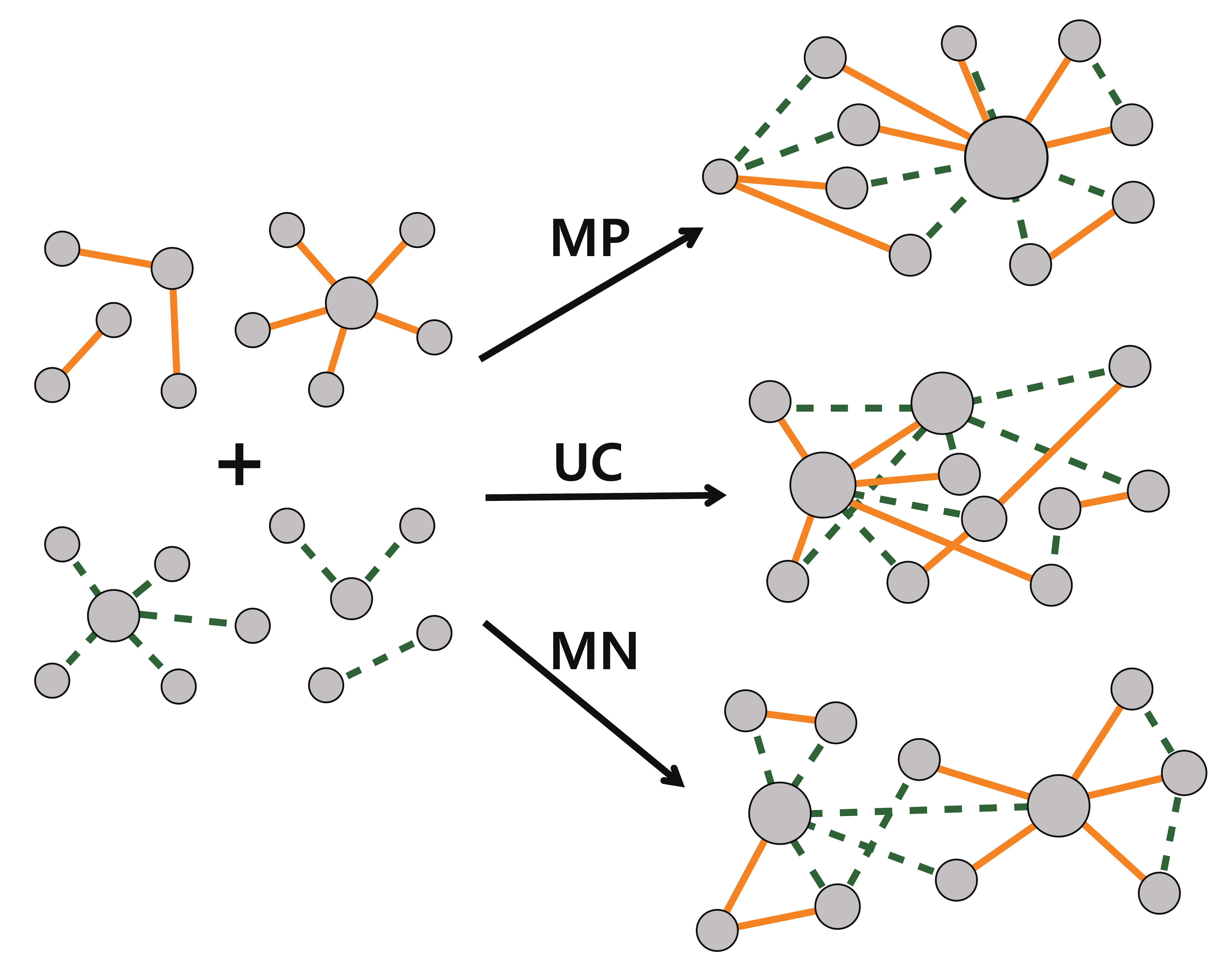}
\caption{
Schematic illustration of three kinds of correlated multiplex networks, maximally-positive (MP), uncorrelated (UC), and maximally-negative (MN).
Each layer of the networks has different types of links, indicated by solid and dashed links, respectively.
}
\end{figure}

To take account of interlayer degree correlations, we mainly consider two layers of multiplex (duplex) networks with comparing three representative correlated structure; maximally-positive (MP), maximally-negative (MN), and uncorrelated (UC) multiplex following Ref.~\cite{kmlee}. In the MP case, node's degrees in different layers are maximally correlated in their degree order, whereas they are maximally anti-correlated in the MN case. Therefore, a node that is the hub in one layer is also the hub in the other layer for the MP case, but it has the smallest degree in the other layer for the MN case. Real-world multiplex networks, of course, would be neither the MP nor the MN case, but the understanding based on these limiting structures with theoretical simplicity can be of illustrative and instructive for building insight towards more realistic situations. 

\section{Biconnectivity}
First, we examine the biconnectivity.  
Subset of nodes in a network connected by at least two disjoint paths is said to form a biconnected
component, or bicomponent for short \cite{bicomponent}.
Existence of the giant bicomponent spanning finite fraction of the entire system is important
for stable connectivity of the network \cite{bicomponent,pkim}.
By definition, all nodes in a bicomponent have at least one alternative way
preserving the connection in networks.
If a typical time scale of the restoration of a broken node is much shorter than that of successive failures,
every node in the bicomponent can completely endure its connectivity.

\subsection{Generating function method}
Generalizing the generating function method from~\cite{leicht} to obtain the size of the giant bicomponent for multiplex networks with $n$ layers, we first define the generating function for the joint degree distribution of $n$ distinct types of links ($n$ layers), $P(\vec{k})$, where $\vec{k}=(k_1,k_2,\cdots,k_n)$ is used to designate the degrees of a node in each layer, as
\begin{eqnarray}
G_0(\vec{x})=\sum_{\vec{k}} P(\vec{k}) \prod_{i=1}^n x_i^{k_i},
\end{eqnarray}
where $\vec{x}=(x_1,x_2,\cdots,x_n)$ is used to denote the auxiliary variables coupled to $\vec{k}$.
We also define the generating function for the remaining degree distribution
by following a randomly chosen $i$-type link, given by 
\begin{eqnarray}
G_1^{(i)}(\vec{x})= \frac{1}{z_i}\frac{\partial}{\partial x_i}G_0(\vec{x}),~
\end{eqnarray}
where $z_i$ is the mean degree of layer $i$.
Then, on locally tree-like networks, the probability $u_i$ that a node reached upon following an $i$-type edge does not belong to the giant component is given by the coupled self-consistency equations
\begin{eqnarray}\label{ui}
u_i=G_1^{(i)}(\vec{u}),
\end{eqnarray}
with $i=1,2,\cdots,n$.
The size of the giant bicomponent, $B$, is equal to the complementary probability
that a randomly chosen node has none or one of its links leading to a node in the giant component \cite{bicomponent},
therefore,
\begin{eqnarray}
B&=&1-G_0(\vec{u})-\sum_i (1-u_i) z_i G_1^{(i)}(\vec{u}),
\end{eqnarray}
where the first two terms give the size of the giant unicomponent, $S=1-G_0(\vec{u})$ \cite{kmlee},
and the last term gives the difference between $S$ and $B$.

\begin{figure}
\includegraphics[width=0.95\linewidth]{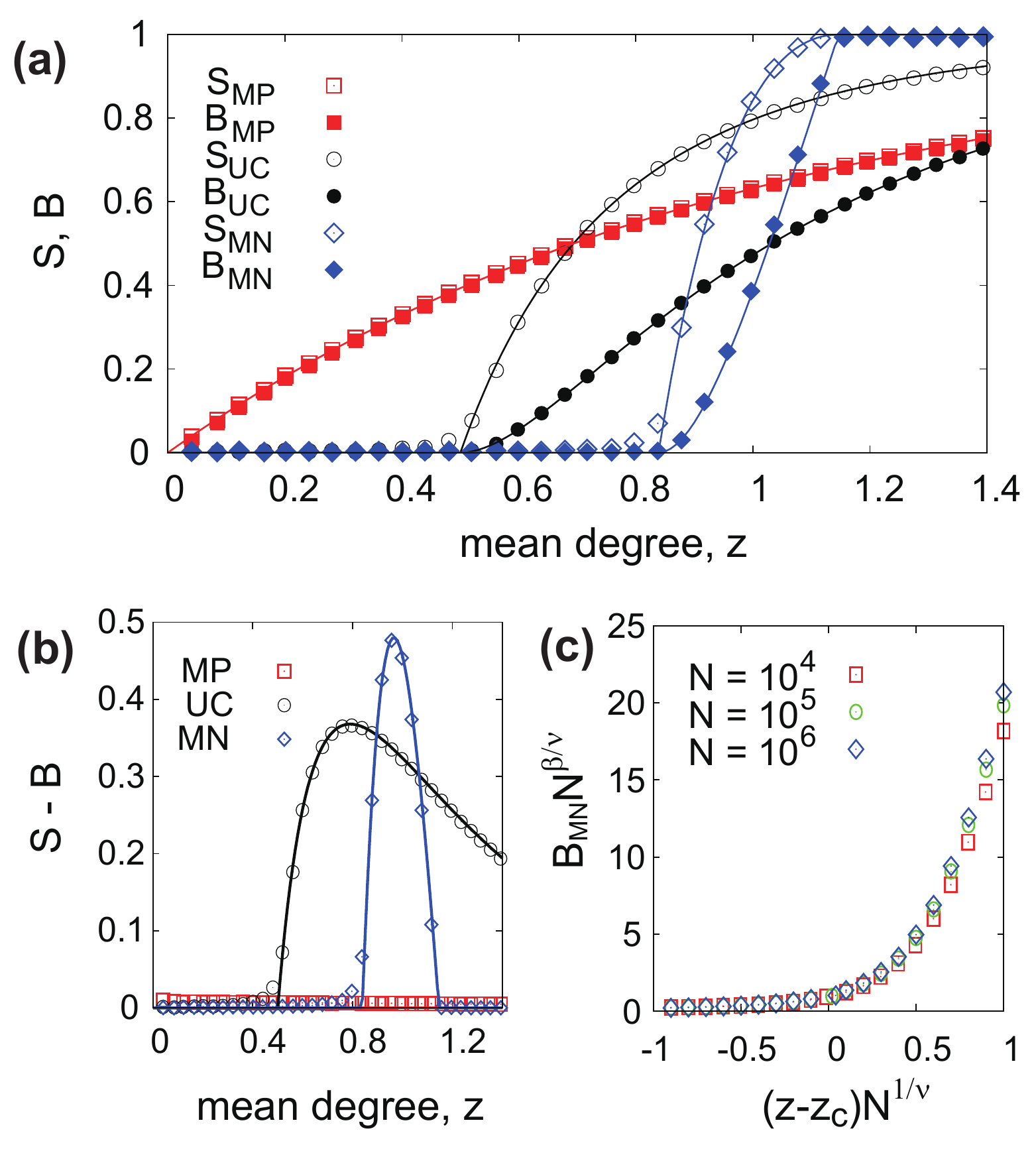}
\caption{
(a) The size of the giant bicomponent, $B$ (filled symbols), and the unicomponent, $S$ (open symbols),
for the MP ($\Box$), the UC ($\circ$), and the MN ($\diamond$) couplings of duplex ER networks.
(b) The gap between $S$ and $B$ as a function of $z$.
Note that $B_{MP}$ is the same with $S_{MP}$.
For the MN coupling, the entire network is connected into a single bicomponent when $z>1.146\dots$.
Theoretical curves (lines) and numerical results (points) obtained with $N = 10^4$ nodes, averaged over $10^3$ runs are shown together.
(c) Data collapse of the scaled bicomponent size for the MN coupling, $B_{MN}N^{\beta/\nu}$, vs. 
the finite-size scaling variable, $(z-z_c)N^{1/\nu}$, with $\beta=2$ and $\nu=3$.
}
\end{figure}

The condition of existence of the giant bicomponent $(B>0)$ is that the largest eigenvalue of 
the Jacobian matrix, $\mathbf{J}$, of Eq.~(\ref{ui}) at $(1,\cdots,1)$ to be larger than $\mathrm{unity}$.
For duplex networks, $\mathbf{J}$ can be expressed as
\begin{eqnarray}
\mathbf{J}=\left( \begin{array}{cc}
\kappa_1 & \mathcal{K}_1 \\
\mathcal{K}_2 & \kappa_2 
\end{array} \right),
\end{eqnarray}
where $\kappa_i=\frac{\langle k_i^2\rangle-z_i}{z_i}$
and $\mathcal{K}_i=\frac{\langle k_1 k_2\rangle}{z_i}$.
The largest eigenvalue $\Lambda$ of $\mathbf{J}$ is given in terms of $\kappa$ and $\mathcal{K}$ as,
\begin{eqnarray}
\Lambda=\frac{1}{2} \left[ \kappa_1+\kappa_2 +\sqrt{(\kappa_1 -\kappa_2)^2+4\mathcal{K}_1\mathcal{K}_2}\right].
\end{eqnarray}

\subsection{Results}
The analytic predictions based on the above generating function method as well as numerical simulation
results are obtained for the duplex Erd\H{o}s-R\'enyi (ER) networks. 
The main results from comparisons of the three correlation types are as follows.
First, the more correlated-coupling there is in multiplex networks, the lower does the percolation threshold become.
Furthermore, the size of the giant bicomponent for the MP case, $B_{MP}$, is the same as that of the giant unicomponent, $S_{MP}$ (Figs.~2a,b), meaning that all pairs of nodes in the giant unicomponent have at least two independently connected paths. In addition, the giant bicomponent always exists for any non-zero link density, so that the MP coupling offers a well-connected structure even with sparse link density.

On the contrary, the emergence of the giant bicomponent for the MN coupling is much delayed.
After passing the percolation threshold, $z_c^{MN}=0.838\dots$,
the size of the bicomponent $B_{MN}$ increases slower than $S_{MN}$ (Fig.~2a,b). 
Near the critical point $z_c^{MN}$, $B_{MN}\sim (z-z_c)^{\beta_B}$, where $\beta_B=2$ (Fig.~2c), 
which is twice the mean-field critical exponent for $S$
in agreement with general critical behavior of bicomponent~\cite{pkim}.
Therefore $B_{UC,~MN}$ increases from zero in a convex manner near $z_c$, in contrast to the behavior of $S$ displaying a concave increase above $z_c$ with $\beta_S=1$ for all three cases \cite{kmlee}.
When $z>z^*=1.146\dots$, the entire network is connected into a single component for the MN coupling and the disparity between $B_{MN}$ and $S_{MN}$ disappears, too.
The maximum value of $(S-B)$ for the MN coupling is located at $z_m^{MN}=0.965\dots$, 
which is larger than that for the UC coupling $z_m^{UC}=0.791\dots$.
The MN coupling hinders the emergence of the giant bicomponent for low density, yet it can establish the biconnected structure over the whole network with a finite link density.

\section{Error and attack tolerance}
The error and attack tolerance of a network under structural disturbance has been one of the major problems in network theory~\cite{albert,callaway,cohen}, which has also been addressed in the context of interdependent networks \cite{buldyrev,parshani-pnas,schneider2,huang} in recent years. In this section, we consider this problem for multiplex networks with interlayer degree correlations. 

\subsection{Generating function method}
For the analytic calculation of the giant component size after removing a fraction of nodes, we extend the generating function method for single networks \cite{callaway} to multiplex networks.
First, let $\phi(\vec{k})$ be the probability that a node with degrees $\vec{k}$ is removed from the initial network, which encodes the node removal strategy.
For example, when $f$ fraction of nodes are removed uniformly by chance, $\phi(\vec{k})=f$.
For the intentional attack in which one removes targeted nodes in order of 
the total degree $K\equiv\sum_{i=1}^n k_i$, one has $\phi(\vec{k})=\Theta(K-K_c)$,
where $\Theta(x)$ is the Heaviside step function and 
$K_c$ is the cutoff total degree for the attack.
With $\phi(\vec{k})$, we can define the joint degree generating function after the node removal as 
\begin{eqnarray}
H_0(\vec{x})=\sum_{\vec{k}} P(\vec{k}) \left[1-\phi(\vec{k}) \right] \prod_{i=1}^n x_i^{k_i}.
\end{eqnarray}
Similarly, the generating function for the remaining degrees upon following a randomly chosen $i$-type link is given by
\begin{eqnarray}
H_1^{(i)}(\vec{x})=\frac{1}{z_i}\frac{\partial}{\partial x_i}H_0(\vec{x}).
\end{eqnarray}
Then, on locally tree-like networks, the probability that a node reached by following an $i$-type link does
not belong to the giant component, $v_i$, is given by the coupled self-consistency equations,
\begin{eqnarray}
v_i=1-H_1^{(i)}(1)+H_1^{(i)}(\vec{v}).
\end{eqnarray}
We finally obtain the giant component size $S$ after the node removal as
\begin{eqnarray}
S=H_0(1)-H_0(\vec{v}),
\end{eqnarray}
with the appropriately chosen $\phi(\vec{k})$ for, e.g., the random breakdown or the intentional attack based on the total degree.
In what follows we present the main results from the analytic calculations together with the numerical simulations on various node removal scenarios and multiplex network couplings. In the first two following subsections, we will demonstrate our analyses on duplex ER networks with layers of equal link density (denoted as $z$), after which the results on other graph ensembles and coupling types are briefly outlined.

\begin{figure}
\includegraphics[width=\linewidth]{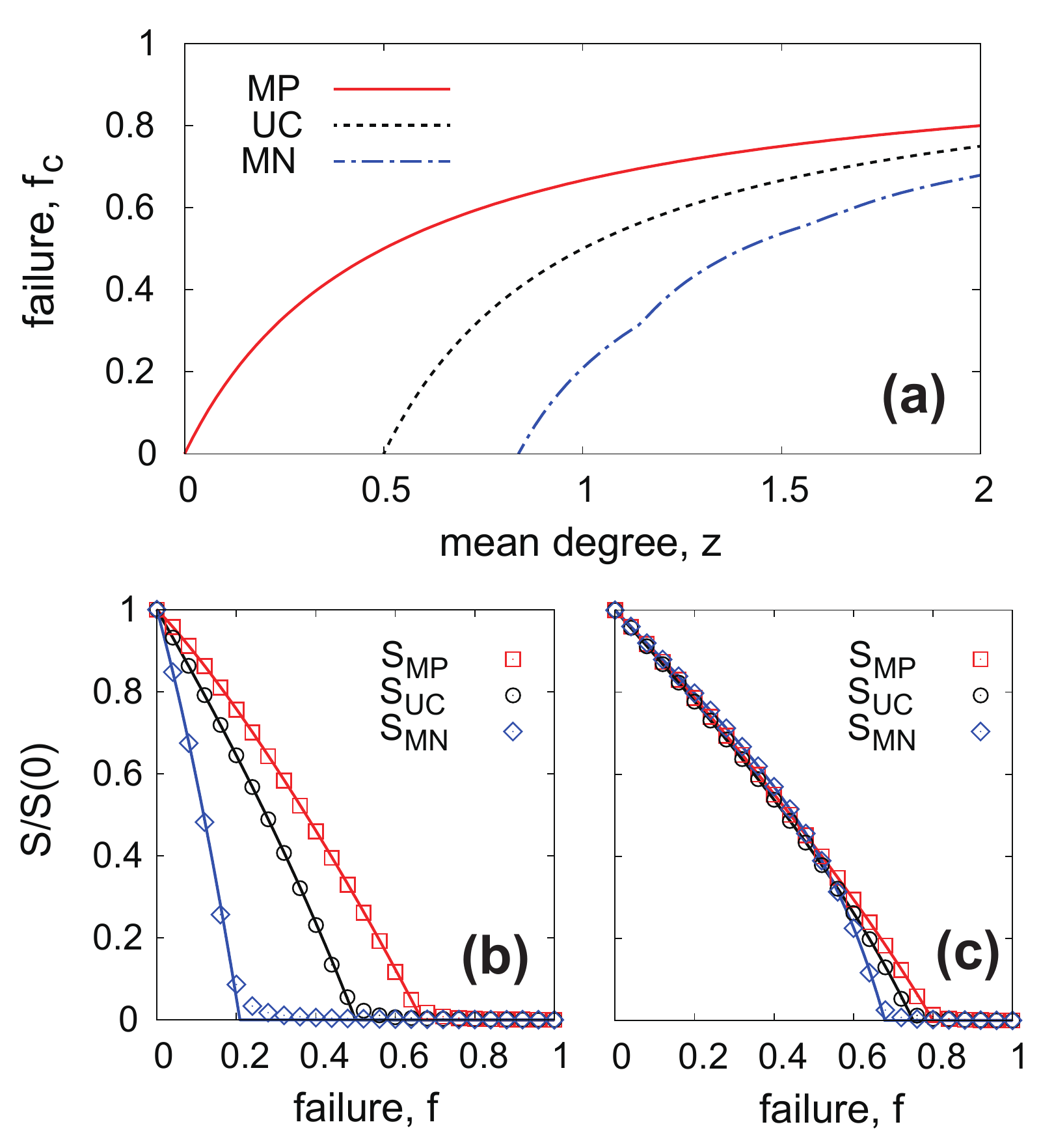}
\caption{
The critical failure fraction (a) and the size of the giant component of 
the correlated duplex ER networks with $z=1$ (b) and $2$ (c)
under random damage.
The MP coupling produces more robust structure than the others against random failure.
Theoretical curves (lines) and numerical results (points) obtained with $N = 10^4$ nodes, averaged over $10^3$ runs are shown together.
}
\end{figure}

\subsection{Error tolerance: Random node removals}
For the random deletion of nodes, that is with $\phi(k_1,k_2)=f$ for duplex networks, the MP (MN) coupling is more resilient (vulnerable) than the others.The percolation threshold for the MP, $f_c^{MP}$, is always larger than that for the UC and the MN couplings, so that more removal of nodes is needed to destroy connection at a given $z$ (Figs.~3a,b). The curve for MN coupling exhibits several kinks, which were found to occur when the minimum total degree of the network changes. Rescaled size of the giant component, $S/S(0)$ where $S(0)$ is the
size of the giant component with $f=0$, for the MP coupling is also larger than those for the other cases for any $f$. Main reason of the high robustness of the MP coupling might be
the skewness of its total degree distribution. By the opposite reason, the MN coupling is more vulnerable under random breakdowns of nodes compared to the UC and the MP cases. Generically the interlayer degree correlation increases the network robustness to random damage, but the effect of correlated multiplexity becomes less significant as the network becomes dense (Fig.~3c).

\begin{figure}
\includegraphics[width=\linewidth]{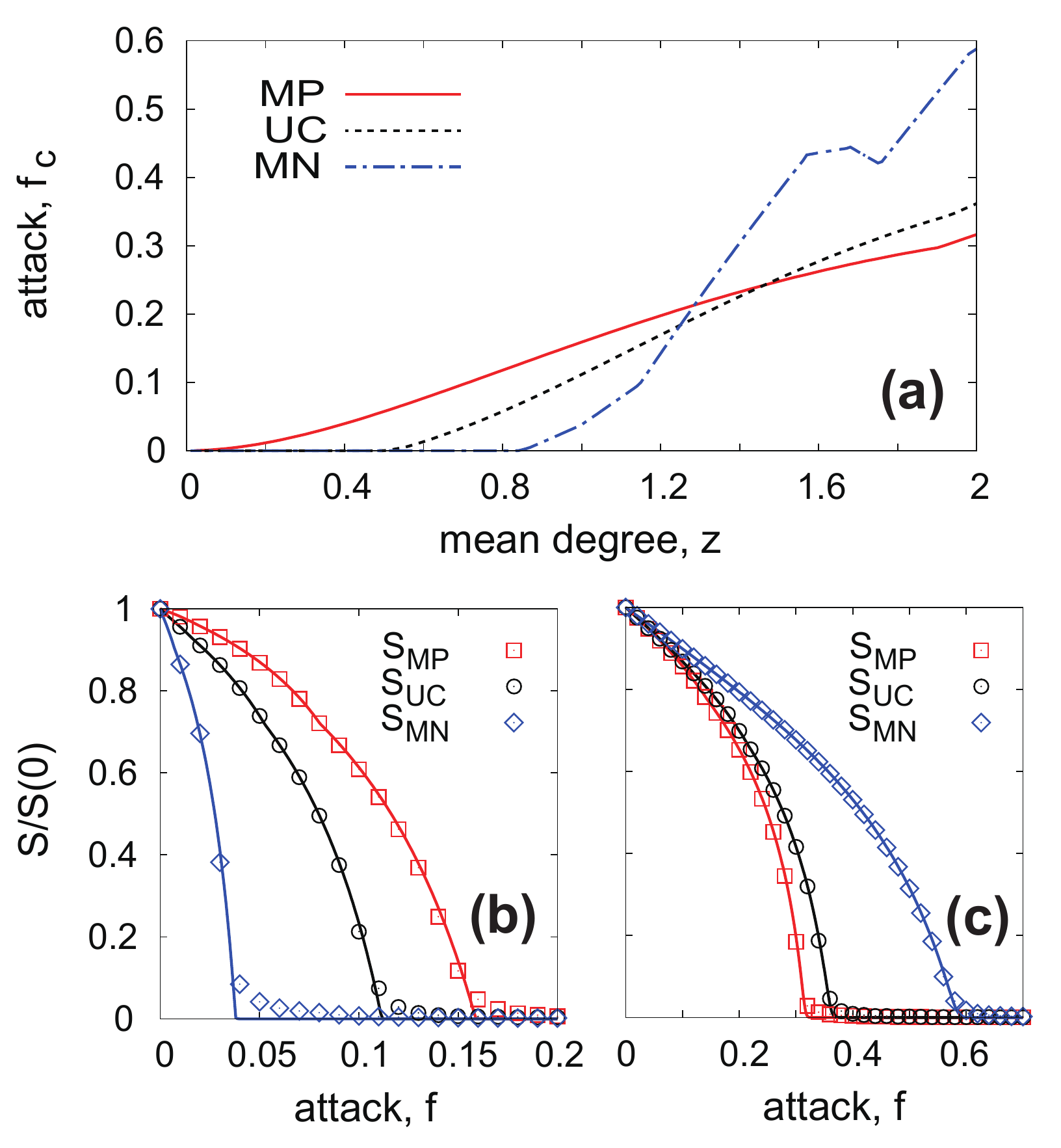}
\caption{
The critical attack fraction (a) and the size of the giant component of 
the correlated duplex ER networks with $z=1$ (b) and $2$ (c)
under the intentional attack based on total degrees.
The MN case is more robust for the dense networks but vulnerable for the sparse networks.
Theoretical curves (lines) and numerical results (points)
obtained with $N = 10^4$ nodes, averaged over $10^3$ runs are shown together.}
\end{figure}

\subsection{Attack vulnerability: Targeted node removals}
For the intentional attack on nodes in the descending order of total degrees, {\it i.e.}, $\phi(k_1,k_2)=\Theta[k_1+k_2-K_c(f)]$ for duplex networks, 
the structural robustness of correlated multiplex networks depends on both the coupling types and link densities, 
as illustrated by the behaviors of the critical attack fraction (Fig.~4a).
When the network is sparse, {\it i.e.} $z<z_\alpha = 1.460\dots$, the MP case is more robust 
against the attack than the UC case (Fig.~4b).
On the contrary, when $z>z_\alpha$, the percolation threshold for the MP coupling is larger than the UC case
meaning that the MP is more vulnerable to the attack in this regime (Fig.~4c).
The MN coupling results in the opposite effect to the MP coupling against the attack.
The MN case is more robust for dense networks but vulnerable for sparse networks than the UC case.
Besides these general trends, 
the critical attack fraction versus the mean degree in duplex ER networks exhibits much more complicated pattern compared to that of random failures, including the anomalous decrease of $f_c$ with respect to $z$, albeit in some narrow windows. 
More detailed investigation would be necessary to examine the structural origin of such anomalies.
Meanwhile, it is well known from single network studies \cite{callaway, cohen-att} that 
networks with more skewed degree distribution are more vulnerable under degree-based attacks in general.
In this perspective, it is interesting to note the MP coupling can produce more robust multiplex network system against the attack for sufficiently sparse link density despite skewness.

\begin{figure}
\includegraphics[width=0.96\linewidth]{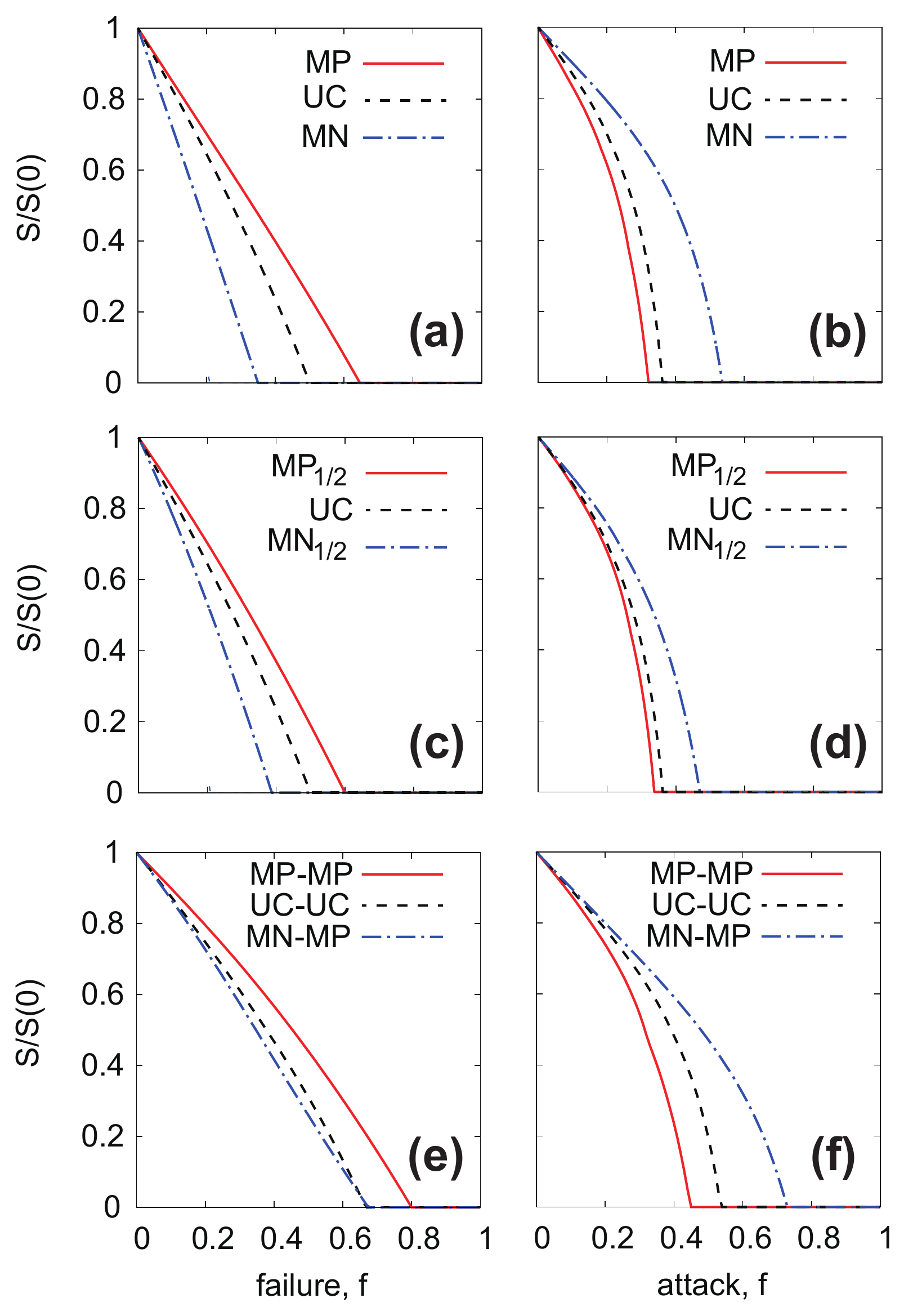}
\caption{
The rescaled size of the giant component on the correlated duplex ER networks 
for the random failure with $z_1=1.5$ and $z_2=0.5$ (a) and the attacks with $z_1=3$ and $z_2=1$ (b),
on partially-correlated duplex ER networks for the failure with $z=1$ (c) and the attack with $z=2$ (d), 
and on triplex ER networks under the failure with $z=1$ (e) and the attack with $z=2$ (f).
}
\end{figure}

\subsection{Other multiplex coupling factors}
To take a more comprehensive overview of the effect of various multiplex coupling factors, 
we consider additional layer coupling scenarios: 
i) duplex ER networks of layers with  different mean degrees, 
ii) duplex ER networks with non-maximal correlated couplings, and iii) triplex ER networks.

First, we examine the duplex ER networks with layers of different mean degree, $z_1 \ne z_2$.
As a specific example, we study the case for $z_1=3 z_2$ against the random failure (Fig.~5a) and the attack (Fig.~5b). The results are qualitatively the same as the equal mean degree case with the same total mean degree: For random failure, the MP coupling is most robust and the MN coupling is least robust (Fig.~5a, to be compared with Fig.~3b). The opposite behaviors are obtained for the targeted attack with $z_1=3z_2=3$ (Fig.~5b, to be compared with Fig.~4c). For $z_1=3 z_2$, the MP case becomes more vulnerable than the UC case against the attack when the total mean degree exceeds $(z_1+z_2)_{\alpha}= 2.522\dots$, which is less than that for the identical mean degree case, suggesting that the layer degree disparity can shrink the regime where the MP coupling is most robust to the attack .

Second, the duplex ER networks with non-maximal correlated coupling are considered. We construct a non-maximal correlated coupling in the following way \cite{kmlee}. A fraction $q$ of nodes are maximally correlated-coupled (either MP or MN) while the other $1-q$ fraction is randomly coupled (UC).
The parameter $q$ sets the strength of correlated coupling between multiplex layers.
In this scheme, the joint degree distribution of the duplex network is obtained by 
$P_{q}(k_1,k_2)=q P_{MAX}(k_1,k_2)+(1-q) P_{UC}(k_1,k_2)$, where $MAX$ is either MP or MN,
which can be readily adopted for theoretical calculation.
The results for $q=1/2$ show that the non-maximal correlation can still affect 
the robustness of networks but the magnitude of the effect is smaller than that of the maximally correlated couplings (Fig.~5c,d; to be compared with Fig. 3d, 4c, respectively).

Finally, we briefly address the robustness of the correlated triplex ER networks with equal  layer-densities.
As there can be two independent interlayer couplings for triplex networks,
there exist a total of six different combinations of layer couplings. 
Here we show the results for three representative coupling combinations:
MP-MP, UC-UC, and MN-MP couplings. 
For example, the MN-MP coupling may represent the case where the first layer is coupled with 
the second layer by the MN coupling whereas it is coupled with the third layer by the MP coupling.
We found that among these three cases the MP-MP coupling is most robust to random node failure
but can be fragile to targeted attack, 
whereas the MN-MP coupling exhibits the opposite behaviors (Fig.~5e,f). The MN-MN coupling gives same results with the MN-MP coupling in this case. We also observed that the MP-UC (MN-UC) coupling yields intermediate behaviors between MP-MP (MN-MP) and UC-UC couplings: 
$f_c$ as well as $S/S(0)$ for MP-UC (MN-UC) lies between those of MP-MP (MN-MP) and UC-UC couplings.

\subsection{Multiplex scale-free networks} 
We also study the same problem for multiplex scale-free (SF) networks numerically. To build the SF network layers with tunable degree exponent $\gamma$ and mean degree, we use the static model~\cite{static}, where each node $s$ $(s=1, 2, \cdots,N)$ has an endogenous weight $\omega_s$ given by $\omega_s=s^{-\mu}/\sum_{t=1}^N t^{-\mu}$, with $\mu$ being a constant, $0<\mu<1$. 
For each step to construct a network, a pair of nodes, say $s$ and $t$, are chosen independently following the probability $\omega_s$ and $\omega_t$, respectively, and connected unless they are already linked. 
One repeats this step until the layer has the desired mean degree $z$. For typical cases with $z={\cal O}(1)$, 
the degree distribution of the resulting layer is asymptotically scale-free, decaying as $\sim k^{-\gamma}$ with the degree exponent $\gamma=(\mu+1)/\mu$.

\begin{figure}
\includegraphics[width=0.85\linewidth]{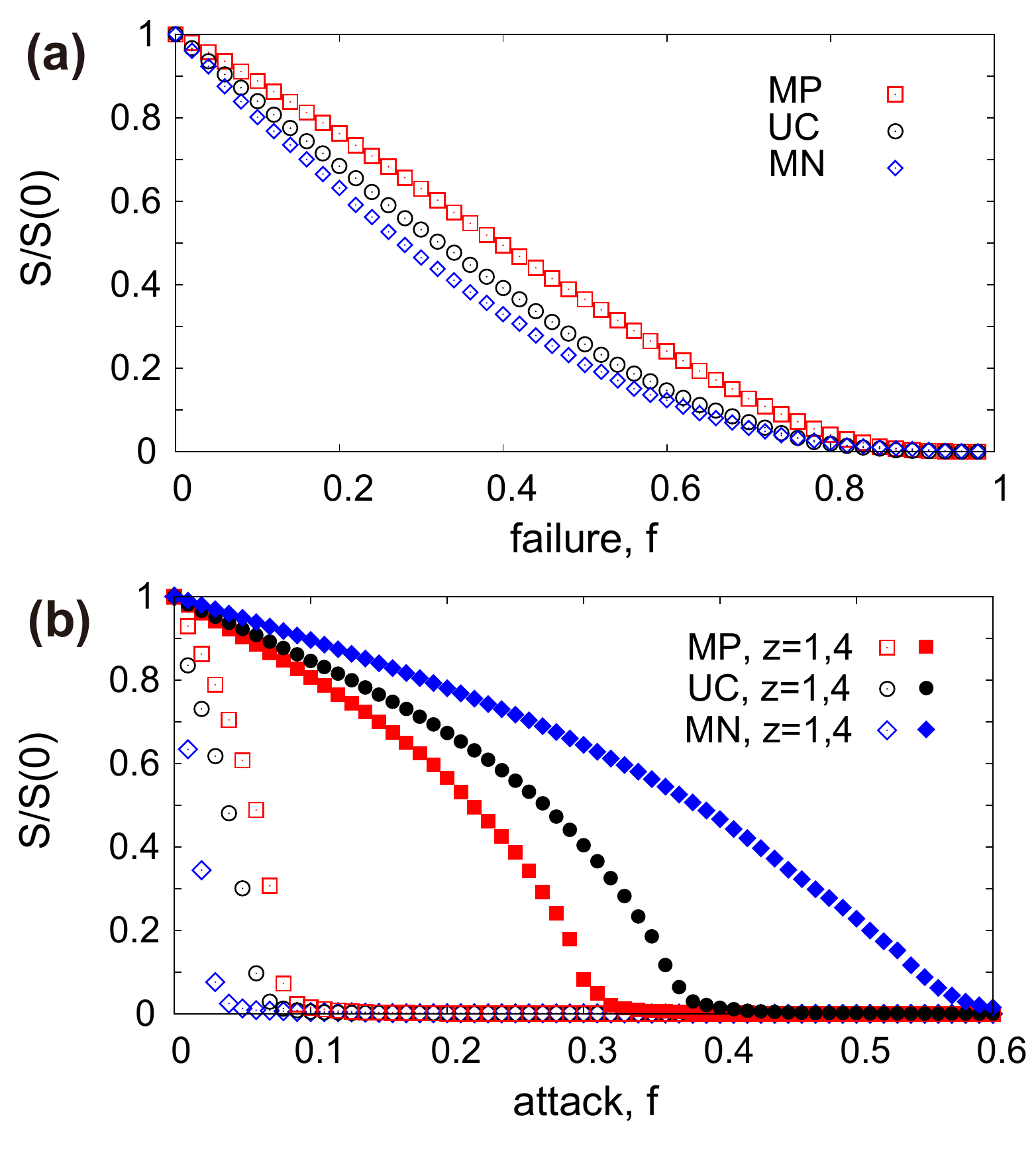}
\caption{
The size of the giant component on the duplex static SF networks with degree exponent $\gamma=2.5$ to the random failure for $z=1$ (a) and the intentional attack based on the total degree for $z=1$ and $4$ (b), obtained with $N = 10^4$ nodes, averaged over $10^3$ runs.
}
\end{figure}

We use SF layers with identical degree exponent $\gamma=5/2$, which is in the range $\gamma\le3$.
In this regime, each layer itself is extremely resilient against the random failures due to high degree heterogeneity, as is well-known from the single-network studies \cite{cohen}. Therefore, all three coupling types show high robustness, with only a small difference among them that the MP coupling is most robust and the MN coupling is least robust, similarly with duplex ER cases (Fig.~6a). For the attack, the MN case is more resilient for dense networks but more vulnerable for sparse networks, again in qualitative similarity to duplex ER cases, as illustrated by the
comparisons of duplex SF networks of equal mean degrees $z=1$ and $4$, respectively~(Fig.~6b). 

\section{Mutual connectivity}

\subsection{Mutual percolation}
In multiplex network systems, layers may be interdependent \cite{buldyrev}, in the sense that nodes in one layer may require supports from corresponding nodes in the other layers and vice versa, demanding simultaneous connectivities in each and every layers of the network for proper function.
For such systems, one can address the network robustness in terms of 
mutually-connected component \cite{buldyrev}, also called mutual component for short,
whose size can be obtained by the generating function method due to Ref.~\cite{swson}, as follows.
On locally tree-like networks, the probability that a node reached by following an $i$-type link
does not belong to the giant mutual component, $w_i$, is given 
by the following coupled self-consistency equations,
\begin{eqnarray}
w_i=1-\sum_{\vec{k}} \frac{k_i P(\vec{k})}{z_i} (1-w_i^{k_i-1}) \prod_{j=1,j\ne i}^{n} (1-w_j^{k_j}).~~
\end{eqnarray}
Then the size of the giant mutual component, $M$, for multiplex networks 
is obtained by
\begin{eqnarray}
M=\sum_{\vec{k}} P(\vec{k}) \prod_{j=1}^{n} (1-w_j^{k_j}).
\end{eqnarray}

\begin{figure}
\includegraphics[width=0.95\linewidth]{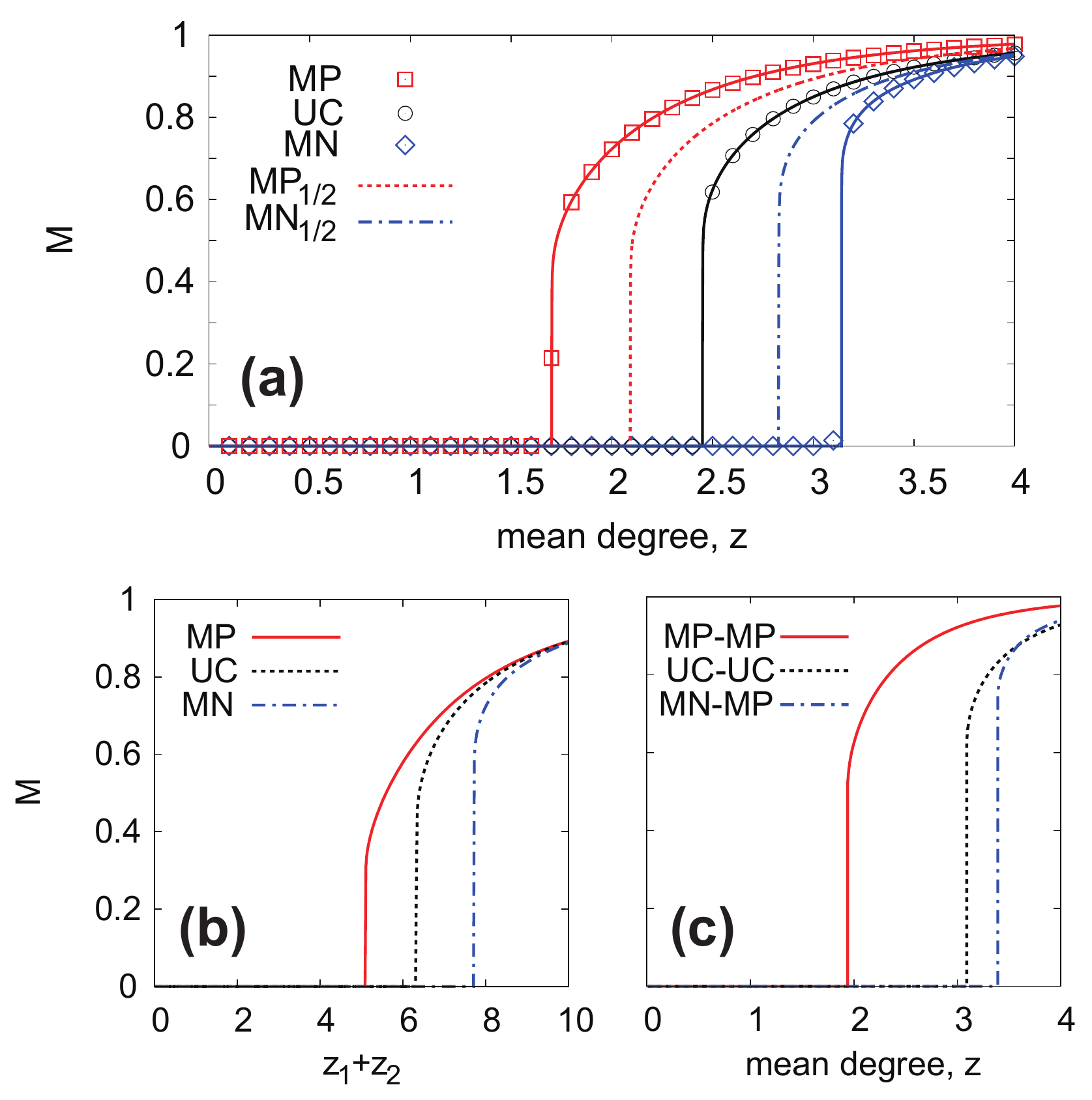}
\caption{
The size of the giant mutual component on the correlated duplex ER networks
with the same mean degree (a) and different mean degrees $z_1=3 z_2$ (b), 
and on the correlated triplex ER networks (c). Lines represent analytical calculations and the symbols in (a) are numerical results obtained with $N = 10^4$ nodes, averaged over $10^3$ runs.
}
\end{figure}

The main results of analytic predictions from the above theory as well as the numerical simulations for the duplex ER networks are as follows (Fig.~7a). As is well-known~\cite{buldyrev,swson}, the giant mutual component emerges discontinuously, in contrast with the ordinary percolation transition that exhibits a continuous phase transition. Similarly to the ordinary connectivity \cite{kmlee}, the percolation threshold of the mutual percolation for the MP coupling is lower, whereas the MN coupling requires denser network for the emergence of the giant mutual component than the other cases. We performed additional analyses on multiplex ER networks, shown in Fig.~7, for the cases of non-maximal correlated couplings (Fig.~7a), unequal layer-densities (Fig.~7b),
and triplex layers (Fig.~7c).

\begin{figure}[t]
\includegraphics[width=0.95\linewidth]{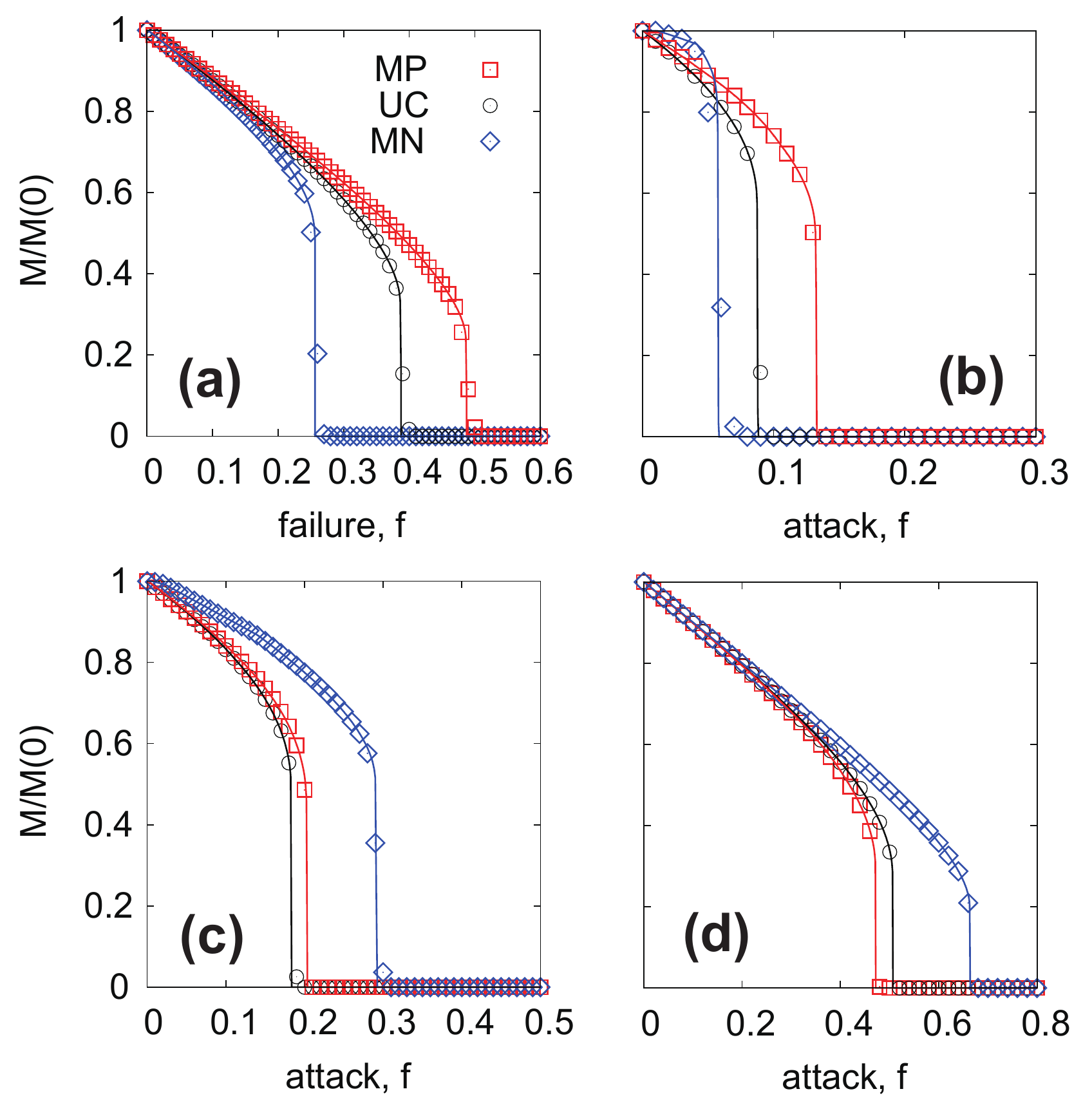}
\caption{
The rescaled size of the giant mutual component on the correlated ER networks
for random failure with $z=4$ (a), and for the attacks based on the degree with $z=3.2$ (b), $4$ (c), and $8$ (d).
Theoretical curves (lines) and numerical results (points)
obtained with $N = 10^4$ nodes, averaged over $10^3$ runs, are shown together.
}
\end{figure}

\subsection{Mutual connectivity under node removals}
Following a similar procedure to the preceding section, one can calculate the giant mutual 
component size under removal of randomly chosen nodes or targeted nodes with the highest degrees 
on locally tree-like networks. Combining the theory for the mutual percolation and the node removals \cite{callaway,swson} (see also \cite{huang} for an alternative approach),
the probability $y_i$ that a node reached by following an $i$-type link
does not belong to the giant mutual component after deletion of nodes
can be obtained by the following coupled self-consistency equations
\begin{align}
y_i=1-\sum_{\vec{k}} \frac{k_i P(\vec{k})}{z_i} \left[1-\phi(\vec{k})\right] (1-y_i^{k_i-1})\prod_{j=1,j\ne i}^{n} (1-y_j^{k_j}).
\end{align}
Then the size of the giant mutual component $M$ after node removals can be computed as
\begin{eqnarray}
M=\sum_{\vec{k}} P(\vec{k}) \left[1-\phi(\vec{k})\right] \prod_{j=1}^{n} (1-y_j^{k_j}).
\end{eqnarray} 

For the duplex ER networks with equal layer-densities, we found that the MP (MN) coupling is more robust (vulnerable) than the other cases against the random node removals. 
The result for the MP case was also obtained earlier in Refs.~\cite{parshani,correlation}.
The rescaled size of the giant mutual component, $M/M(0)$ where $M(0)$ is the size of the giant mutual component with $f=0$, for the MP (MN) coupling is larger (smaller) than those for the others for any removal fraction $f$ (Fig.~8a). For the targeted attack based on the total degree, however, the effect of correlated multiplexity is more complicated. For sufficiently low density, e.g., $z\approx3.2$ (Fig.~8b), the MP (MN) coupling is more robust (vulnerable) than the others against the attack. With intermediate density, say $z\approx 4$ (Fig.~8c), the MN coupling is most robust and the UC is most vulnerable. For high enough density, e.g., $z\approx 8$ (Fig.~8d), the MN (MP) case is most robust (vulnerable) against the attack, opposite to the low density case. This shows that the effect of correlated multiplexity on the robustness of mutual connectivity is not monotonic and could depend strongly on the details of interdependency.

\section{A real-world example}
Finally, we examine the robustness property of a real-world multiplex network under node removals. The real-world network data we consider consists of two layers, the Internet backbone network and the high-voltage electrical transmission network in Italy~\cite{rosato}. These two network layers can be regarded as interdependent in such a way that a failure in one layer (say, a power station in the power grid) would lead to that on the other layer (say, a power control station communicating through the Internet), and vice versa. Thus this system can be modeled as a multiplex network. Following the rationale of \cite{buldyrev,parshani}, 
we have established the interdependency between two layers based on the geographical distance so that each node in the Internet network is interdependent on the closest node in the power transmission network. Nodes with no interdependent partner are thought to be functional autonomously.

\begin{figure}[t]
\includegraphics[width=0.99\linewidth]{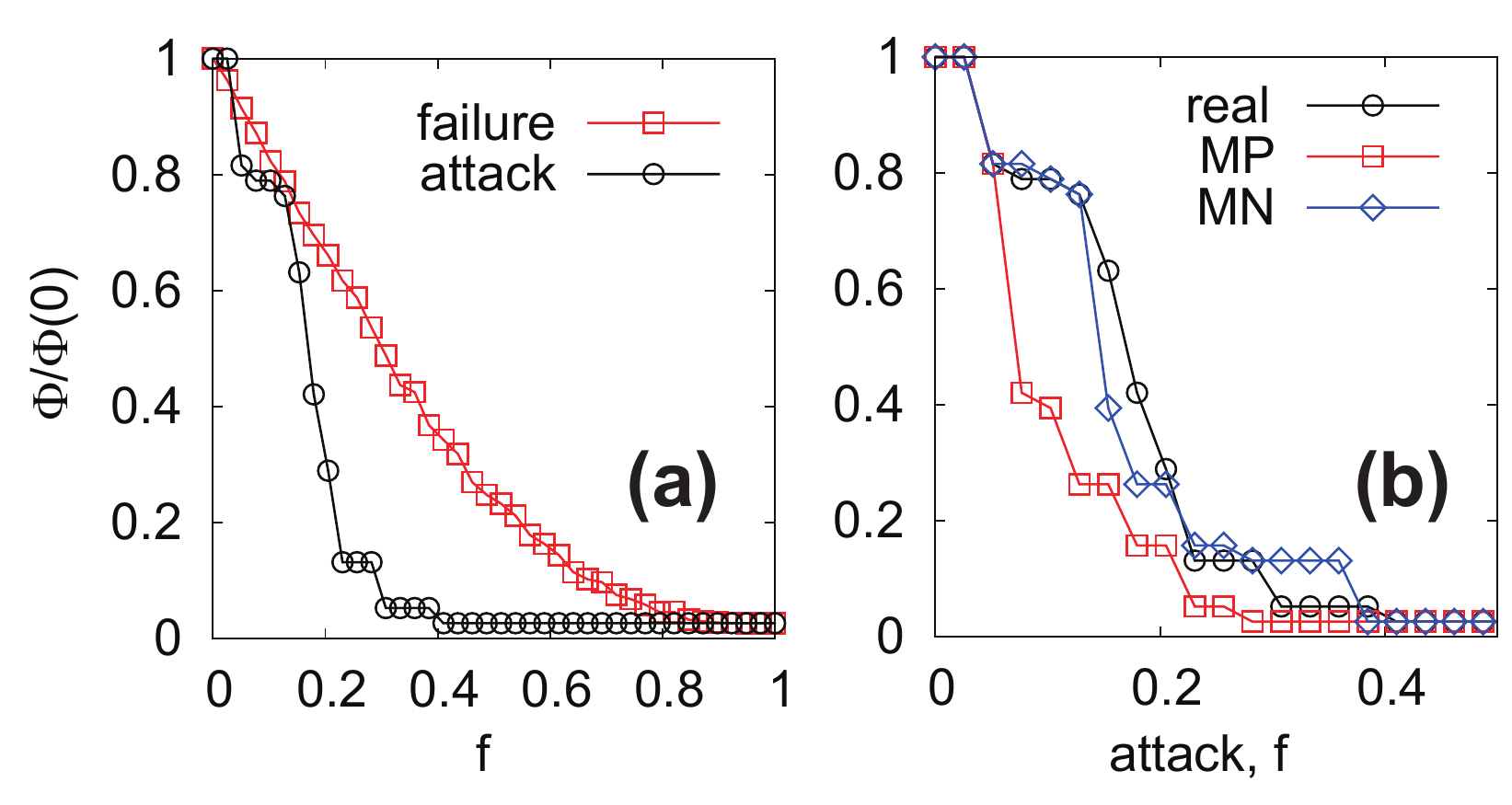}
\caption{
(a) Rescaled size of functional interdependent nodes against random node failure and targeted attack,
simulated for the real-world network data from the Italian Internet-power transmission multiplex network \cite{rosato}. (b) Same plots for targeted attack on the rewired networks. Symbols are numerical simulation results, connected by guidelines for visibility.}
\end{figure}

We first calculate numerically the fraction of functional interdependent nodes, $\Phi$,  
of the Internet-power transmission multiplex network following the interdependent cascade model of Ref.~\cite{parshani2} upon the random failure and the degree-based targeted attack on the interdependent nodes (Fig.~9a). The numerical results show that the rescaled fraction of functional nodes,  $\Phi/\Phi(0)$ where $\Phi(0)$ is the fraction of functional nodes with $f=0$, is relatively robust against the random failure as it can endure  up to around 80\% interdependent-node removals, whereas it rapidly disintegrates upon the targeted attack on as small as 20\% of highest-degree interdependent nodes. We also examine the effect of correlated couplings in this system to the attack vulnerability by using artificial multiplex networks with rewired interdependency into the MP or the MN types (Fig.~9b). The results for the rewired multiplex networks show that 
the MN coupling is more robust to the targeted attack on high-degree nodes than the MP coupling. It is interesting to note that the behavior of the real-world network data lies close to that of the MN coupling despite significant difference in actual interdependency patterns.

\section{Summary}	
In this paper, we have studied various network robustness properties of multiplex networks
focusing on the role of the correlation between degrees of a node across different layers.
We have analyzed specifically the biconnectivity and 
the error and attack tolerance of the ordinary as well as the mutual connectivity, 
covering a wide spectrum of network robustness relevant to multiplex networks.
We found that the correlated coupling of multiplex layers can significantly alter 
the robustness properties of multiplex networks in diverse ways.
For example, positively-correlated multiplex networks are more robust, whereas
anti-correlated multiplex are less robust, in the context of the biconnectivity and 
the ordinary as well as mutual connectivity upon random node failure.
To the targeted attack based on nodes' degrees, on the contrary, 
positively-correlated multiplex networks with sufficiently high link-density can be 
highly vulnerable, whereas the anti-correlated ones can become more resilient. 
We also examined the effect of various additional multiplex-coupling factors and a real-world example
of the Italian Internet-power transmission multiplex system.

Our analyses reveal that the notion of network robustness can exhibit more diversified aspects in multiplex networks compared to single-network situation, dependent on specific context and interplay between the network layers. 
We expect our initial analyses could prompt attention and provide a basic insight for further research endeavors on understanding the robustness of correlated multiplex systems. 
Interesting topics of future work in this regard would include the extension to account for higer-order correlation properties beyond the interlayer degree correlation considered in this work, such as clustering~\cite{clustering} in multiplex networks.

\begin{acknowledgments}
We thank V. Rosato for providing the Italian Internet backbone
and the high-voltage electrical transmission network data.
We also thank the anonymous referees for useful comments.
This work was supported by Basic Science Research Program through the NRF grant funded by MSIP (No. 2011-0014191). 
B. M. is also supported by a Korea University Grant.
K.-M. L. is also supported by Global Ph.D. Fellowship Program (No.\ 2011-0007174) through NRF, MSIP.
\end{acknowledgments}

\end{document}